\documentclass{sigplanconf}
\usepackage[english]{babel}
\usepackage{blindtext}
\usepackage{listings}

\usepackage{hyperref}
\usepackage{upquote}
\usepackage{color,soul}
\usepackage{graphicx}
\usepackage{multirow}
\usepackage{booktabs}
\usepackage{dcolumn}

\usepackage{caption}\DeclareCaptionType{copyrightbox}

\usepackage{subcaption}

\newcommand{\cs}{C\#}
\newcommand{\fs}{F\#}
\newcommand{\java}{Java}
\newcommand{\scala}{Scala}

\newcommand{\sv}[1]{{\small\texttt{\tt #1}}}

\definecolor{dkgreen}{rgb}{0,0.6,0}
\definecolor{gray}{rgb}{0.5,0.5,0.5}
\definecolor{mauve}{rgb}{0.58,0,0.82}
\definecolor{light-gray}{gray}{0.25}

\lstdefinestyle{java}{
  language=Java,
  aboveskip=3mm,
  belowskip=3mm,
  showstringspaces=false,
  columns=flexible,
  basicstyle={\footnotesize\ttfamily},
  numberstyle={\tiny},
  numbers=left,
  keywordstyle=\color{blue},
  commentstyle=\color{dkgreen},
  stringstyle=\color{mauve},
  breaklines=true,
  breakatwhitespace=true,
  tabsize=3,
  keepspaces
}

\lstdefinestyle{bytecode}{
  otherkeywords={invokedynamic},
  language=JVMIS,
  aboveskip=3mm,
  belowskip=3mm,
  showstringspaces=false,
  columns=flexible,
  basicstyle={\footnotesize\ttfamily},
  numberstyle={\tiny},
  numbers=left,
  keywordstyle=\color{blue},
  commentstyle=\color{dkgreen},
  stringstyle=\color{mauve},
  breaklines=true,
  breakatwhitespace=true,
  tabsize=3,
  keepspaces
}

\lstdefinestyle{fsharp} {	
  morekeywords={let, new, match, with, rec, 
    open, module, namespace, type, of, member, 
    and, for, while, true, false, in, do, begin, 
    end, fun, function, return, yield, try, val, 
    mutable, if, then, else, cloud, async, static, 
    use, abstract, interface, inherit, finally, maybe, option },
  otherkeywords={ let!, return!, do!, yield!, use!, var, from, select, where, order},
  keywordstyle=\color{blue},
  sensitive=true,
  aboveskip=3mm,
  belowskip=3mm,
  showstringspaces=false, 
  keepspaces,
  columns=flexible,
  basicstyle={\footnotesize\ttfamily},
  numberstyle={\tiny},
  numbers=left,
  breaklines=true,
  upquote=true,
  tabsize=3,
  morecomment=[l][\color{dkgreen}]{///},
  morecomment=[l][\color{dkgreen}]{//},
  morecomment=[s][\color{dkgreen}]{{(*}{*)}},
  morestring=[b]",
  showstringspaces=false,
  literate={`}{\`}1,
  stringstyle=\color{mauve}
}

\lstdefinestyle{csharp} {
  aboveskip=3mm,
  belowskip=3mm,
  showstringspaces=false,
  columns=flexible,
  basicstyle={\footnotesize\ttfamily},
  numberstyle={\tiny},
  numbers=left,
  keywordstyle=\color{blue},
  commentstyle=\color{dkgreen},
  stringstyle=\color{mauve},
  breaklines=true,
  breakatwhitespace=true,
  tabsize=3,
  morecomment = [l]{//}, 
  morecomment = [l]{///},
  morecomment = [s]{/*}{*/},
  morestring=[b]", 
  keepspaces,
  sensitive = true,
  morekeywords = {async, await, abstract,  
    event,  new,  struct,
    as,  explicit,  null,  switch,
    base,  extern,  object,  this,
    bool,  false,  operator,  throw,
    break,  finally,  out,  true,
    byte,  fixed,  override,  try,
    case,  float,  params,  typeof,
    catch,  for,  private,  uint,
    char,  foreach,  protected,  ulong,
    checked,  goto,  public,  unchecked,
    class,  if,  readonly,  unsafe,
    const,  implicit,  ref,  ushort,
    continue,  in,  return,  using,
    decimal,  int,  sbyte,  virtual,
    default,  interface,  sealed,  volatile,
    delegate,  internal,  short,  void,
    do,  is,  sizeof,  while,
    double,  lock,  stackalloc,   
    else,  long,  static,   
    enum,  namespace,  string, from, select, where}
}

\lstdefinestyle{scala} {  
  morekeywords={ abstract,case,catch,
    char,class,
    def,else,extends,final,
    if,import,
    match,module,new,null,object,
    override,package,private,protected,
    public,return,super,this,throw,
    trait,try,type,val,var,with,implicit,
    macro,sealed
  },
  sensitive,
  morecomment=[l]//,
  morecomment=[s]{/*}{*/},
  morestring=[b]",
  morestring=[b]',
  aboveskip=3mm,
  belowskip=3mm,
  showstringspaces=false,
  columns=flexible,
  basicstyle={\footnotesize\ttfamily},
  numberstyle={\tiny},
  numbers=left,
  keywordstyle=\color{blue},
  commentstyle=\color{dkgreen},
  stringstyle=\color{mauve},
  breaklines=true,
  breakatwhitespace=true,
  tabsize=3,
  keepspaces
}

\DeclareCaptionFormat{myformat}{#1#2#3\hrulefill}
\begin{document}

\setlength{\pdfpageheight}{\paperheight}
\setlength{\pdfpagewidth}{\paperwidth}

\makeatletter
\def\@copyrightspace{\relax}
\makeatother

\title{Clash of the Lambdas}
\subtitle{Through the Lens of Streaming APIs}

\authorinfo{Aggelos Biboudis}
  {University of Athens}
  {biboudis@di.uoa.gr}
\authorinfo{Nick Palladinos}
  {Nessos Information Technologies, SA}
  {npal@nessos.gr}
\authorinfo{Yannis Smaragdakis}
  {University of Athens}
  {smaragd@di.uoa.gr}

\maketitle

\begin{abstract}

The introduction of lambdas in Java 8 completes the slate of
statically-typed, mainstream languages with both object-oriented and
functional features. The main motivation for lambdas in Java has been
to facilitate stream-based declarative APIs, and, therefore, easier
parallelism. In this paper, we evaluate the performance impact of
lambda abstraction employed in stream processing, for a variety of
high-level languages that run on a virtual machine 
(\cs{}, \fs{}, \java{} and \scala{}) and runtime
platforms (JVM on Linux and Windows, .NET CLR for Windows, Mono for
Linux). Furthermore, we evaluate the performance gain that two
optimizing libraries (\mbox{ScalaBlitz} and LinqOptimizer) can offer for
\cs{}, \fs{} and \scala{}. Our study is based on small-scale
throughput-benchmarking, with significant care to isolate different
factors, consult experts on the systems involved, and identify causes
and opportunities. We find that Java exhibits high implementation
maturity, which is a dominant factor in benchmarks. At the same time,
optimizing frameworks can be highly effective for common query patterns.

\end{abstract}


% general terms are not compulsory anymore, 
% you may leave them out


\section{Introduction}

Java 8 has introduced lambdas with the explicit purpose of enabling
streaming abstractions. Such abstractions present an accessible,
natural path to multicore parallelism---perhaps the highest value
domain in current computing. Other languages, such as \scala{}, \cs{},
and \fs{}, have supported lambda abstractions and streaming APIs,
making them a central theme of their approach to parallelism.
Although the specifics of each API differ, there is a core of common
features and near-identical best-practices for users of these APIs in
different languages.

Streaming APIs allow the high-level manipulation of value streams
(with each language employing slightly different terminology) with
functional-inspired operators, such as \sv{filter}, or \sv{map}.  Such
operators take user-defined functions as input, specified via local
functions (lambdas). The Java example fragment below shows a ``sum of
even squares'' computation, where the even numbers in a sequence are
squared and summed. The input to \sv{map} is a lambda, taking an
argument and returning its square. This particular lambda application
is \emph{non-capturing}: the bodies of the lambda expressions in lines 3,4 use only
their argument values, and no values from the environment.

\begin{minipage}{\linewidth}\begin{lstlisting}[style=Java]
public int sumOfSquaresEvenSeq(int[] v) {
  int sum = IntStream.of(v)
    .filter(x -> x % 2 == 0)
    .map(x -> x * x)
    .sum();
  return sum;
}
\end{lstlisting}\end{minipage}

The above computation can be trivially parallelized with the addition
of a \sv{.parallel()} operator before the call to \sv{filter}. This
ability showcases the simplicity benefits of streaming abstractions
for parallel operations.

In this paper, we perform a comparative study of the lambda+streams
APIs of four multi-paradigm, virtual machine-based languages, 
\java{}, \scala{}, \cs{}, and \fs{}, with an emphasis on
implementation and performance comparison, across mainstream platforms
(JVM on Linux and Windows, .NET CLR for Windows, Mono for Linux). We
perform micro-benchmarking\footnote{Code in
  https://github.com/biboudis/clashofthelambdas .} and aim to get a
high-level understanding of the costs and causes. Our goal is the usual goal of
microbenchmarking: to minimize most threats-to-validity by controlling
external factors. (The inherent drawback of microbenchmarking, which
we do not attempt to address, is the threat that benchmarks are not
representative of real uses.) In order to control external factors, we
attempt to select equivalent abstractions in all settings, isolate
dependencies, employ best-practice benchmarking techniques, and
repeatedly consult experts on the different platforms.

Since lambdas+stream operators have arisen independently in so many
contexts and have been central in parallel programming strategies, one
would expect them to be well-understood: a mainstream, high-value
feature is expected to have fairly uniform implementation techniques
and trade-offs. Instead we find interesting variation, even in the
compilation to intermediate code (per platform, e.g., across Java and
Scala, which are both JVM languages). Furthermore, we find that JIT
optimization inside the VM does not always interact predictably with
the code produced for lambdas. This was a minor surprise, given the
maturity of the respective facilities.\footnote{Although Java lambdas
  are standard only as of version 8, their arrival had been
  forthcoming since at least 2006.}

A second aspect of declarative streaming operations is that they
enable aggressive optimization~\cite{murray_steno:_2011}. Optimization
frameworks, such as
LinqOptimizer~\cite{nick_palladinos_linqoptimizer:_2013} and
ScalaBlitz~\cite{aleksandar_prokopec_scalablitz:_2013,prokopec_lock-free_2014},
recognize common patterns of streaming operations and optimize them,
by inlining, performing loop fusion, and more.

In all, we find that Java offers high performance for lambdas and
streaming operations, primarily due to optimizing for non-capturing
lambdas. At the same time, Java suffers from the lack of an optimizing
framework---LinqOptimizer and ScalaBlitz give a significant boost to
\cs{}/\fs{} and \scala{} implementations, respectively, when
optimizations are applicable.

\section{Implementation Techniques for Lambdas and Streaming}

As part of our investigation, we examined current APIs and
implementation techniques for lambdas and streaming abstractions in
the languages and libraries under study. We detail such elements next,
so that we can refer to them directly in our experimental results.

\subsection{Programming Languages}

We begin with the API and implementation description for the languages
of our study: \java{}, \scala{}, and \cs{}/\fs{} (the latter are
similar enough that are best discussed together, although they exhibit
non-negligible performance differences).

\subsubsection{\java}

\java{} is probably the best reference point for our study, although
it is also the relative newcomer among the lambdas+streaming
facilities. We already saw examples of the \java{} API for streaming
in the Introduction. In terms of implementation, the \java{} language
team has chosen a translation scheme for lambdas that is highly 
optimized and fairly unique among statically typed languages.

In the \java{} 8 declarative stream processing API, operators fall
into two categories: intermediate (\emph{always lazy}---e.g., map and
filter) and terminal (which can produce a value or perform
side-effects---e.g., sum and reduce). For concreteness, let us
consider the pipeline below. The following expression (serving as a
running example in this section) calculates the sum of all values in a
double array.

\noindent\begin{minipage}{\linewidth}\begin{lstlisting}[style=Java]
public double sumOfSquaresSeq(double[] v) {
  double sum = DoubleStream.of(v)
  .map(d -> d * d)
  .sum();
  return sum;
}
\end{lstlisting}\end{minipage}

The code first creates a sequential, ordered \sv{Stream} of
\sv{double}s from an array that holds all values. (\sv{DoubleStream}
represents a primitive specialization of \sv{Stream}--one of three
specialized \sv{Stream}s, together with \sv{IntStream} and
\sv{LongStream}.) The calls \sv{map} and \sv{sum} are an intermediate
and a terminal operation respectively. The first operation returns a
\sv{Stream} and it is lazy. It simply declares the transformation that
will occur when the stream will be traversed. This transformation is a
stateless operation and is declared using a (non-capturing) lambda
function. The second operation needs all the stream processed up to
this point, in order to produce a value; this operation is \sv{eager}
and it is effectively the same as reducing the stream with the lambda
\sv{(x,y) -> x+y}.

Implementation-wise, the (stateless or stateful) operations on a
stream are represented by objects chained together sequentially. 
A terminal operation triggers the evaluation of the chain.  In
our example, \emph{if no optimization were to take place}, the
\sv{sum} operator would retrieve data from the stream produced by
\sv{map}, with the latter being supplied the necessary lambda
expression. This traversing of the elements of a stream is realized
through \sv{Spliterator}s. The \sv{Spliterator} interface offers an API for
traversing and partitioning elements of a source and it can operate either
sequentially or in parallel. \sv{Spliterator}s are also equiped with more
advanced functionality---e.g., they can detect structural interference with the
source while processing. The definition of a stream and operations on it are
usually described declaratively and the user does not need to invoke operations
on a \sv{Spliterator} (although a controlled traversal is possible). The \sv{Spliterator}
interface is shown below.

\noindent\begin{minipage}{\linewidth}\begin{lstlisting}[style=java]
public interface Spliterator<T> {
  boolean tryAdvance(Consumer<? super T> action);
  void forEachRemaining(Consumer<? super T> action);
  Spliterator<T> trySplit();
  long estimateSize();
  long getExactSizeIfKnown();
  int characteristics();
  boolean hasCharacteristics(int characteristics);
  Comparator<? super T> getComparator();
}
\end{lstlisting}\end{minipage}

Normally, for the general case of standard stream processing, the
implementation of the above interface will have a
\sv{forEachRemaining} method that internally calls methods
\sv{hasNext} and \sv{next} to traverse a collection, as well as
\sv{accept} to apply an operation to the current element. Thus, three
virtual calls per element will occur.

However, stream pipelines, such as the one in our example, can be
optimized. For the array-based \sv{Spliterator}, the
\sv{forEachRemaining} method performs an indexed-based, do-while
loop. The entire traversal is then transformed: instead of \sv{sum}
requesting the next element from \sv{map}, the pipeline operates in
the inverse order: \sv{map} pushes elements through the \sv{accept}
method of its downstream \sv{Consumer} object, which implements the
\sv{sum} functionality. In this way, the implementation eliminates two
virtual calls per step of iteration and effectively uses internal
iteration, instead of external.

The following (simplified for exposition) snippet of code is taken
from the \sv{Spliterators.java} source file of the Java 8 library and
it demonstrates this special handling, where \sv{a} holds the source
array and \sv{i} indexes over its length:

\noindent\begin{minipage}{\linewidth}\begin{lstlisting}[style=Java]
do { consumer.accept(a[i]); } while (++i < hi);
\end{lstlisting}\end{minipage}

The internal iteration can be seen in this code. Each of the operators
applicable to a stream needs to support this inverted pattern by
supplying an \sv{accept} operation. That operation, in turn, will call
\sv{accept} on whichever \sv{Consumer<T>} may be downstream. For
instance, the fragment of the \sv{map} implementation below shows the
\sv{accept} call (line 7) on the next operator (\sv{sum} in our
example). The code also shows the call to \sv{apply}, invoking the
passed lambda.

\begin{lstlisting}[style=Java]
<T, R> Stream<R> map(Stream<T> source, 
                     Function<T, R> mapper) {
  return new MapperStream<T, R>(source) {
    Consumer<T> wrap(Consumer<R> consumer) {
      return new Consumer<T>() {
        void accept(T v) 
        { consumer.accept(mapper.apply(v)); }
      };
    }
  };
}
\end{lstlisting}

Having seen the implementation of streams, we now turn our attention
to lambdas. There could be several potential translations for lambdas,
such as inner-classes (for both capturing or non-capturing, lambdas),
translation based on \sv{MethodHandle}s---the dynamic and strongly
typed component that was introduced in JSR-292---and more. Each option
has some advantages and disadvantages. For the translation of lambdas
in \java{} 8, the compiler incorporates a technique based on
JSR-292~\cite{rose_jsr_2011} and more specifically on the new
\sv{invokedynamic} command~\cite[Chapter~6]{tim_lindholm_java_2014}
and \sv{MethodHandle}s~\cite{brian_goetz_translation_2012}. 

When the compiler encounters a lambda function, it desugars it to a
method declaration and emits an \sv{invokedynamic} instruction at that
point. For instance, our \sv{sumOfSquaresSeq} example compiles to the
bytecode below:

\noindent\begin{minipage}{\linewidth}\begin{lstlisting}[style=bytecode]
...                         // v on the stack
invokestatic  #7            // DoubleStream.of
invokedynamic #10,  0       // applyAsDouble
invokeinterface #11,  2     // DoubleStream.map
invokeinterface #8,  1      // DoubleStream.sum
dstore_1      
dload_1       
dreturn    
\end{lstlisting}\end{minipage}

Note the \sv{invokedynamic} instruction on line 3, used to return an object that
represents a lambda closure. The method invoked is
\sv{LambdaMetafactory.metafactory}---implemented as part of the Java standard
library. The fully dynamic nature of the call is due to having a single
implementation for retrieving objects for any given method signature. This
process involves three phases: \emph{Linkage}, \emph{Capture} and \emph{Invocation}. When
\sv{invokedynamic} is met for the first time it must link this site with a
method. For the lambda translation case, an instance of \sv{CallSite} is
generated whose target knows how to create function objects. This target
(\sv{LambdaMetafactory.metafactory}) is a factory
for function objects. The Capture phase may involve
allocation of a new object that may capture parameters or will always return the
same object (if no parameters are captured). The third phase is the actual invocation. The advantage of this
translation scheme is that, for lambdas that do not capture any free variables,
a single instance for all usages is enough. Furthermore, the call site is linked
only once for successive invocations of the lambda and, after that, the JVM
inlines the retrieved method's invocation at the dynamic call
site. Additionally, there is no performance burden for loading a class from
disk, as there would be in the case of a fully static translation.

\subsubsection{\scala}

\scala{} is an object-functional programming language for the
JVM. \scala{} has a rich object system offering traits and mixin
composition. As a functional language, it has support for higher-order
functions, pattern matching, algebraic data types, and more. Since
version 2.8, \scala{} comes with a rich collections library offering a
wide range of collection types, together with common functional
combinators, such as \sv{map}, \sv{filter}, \sv{flatMap}, etc. There are
two \scala{} alternatives for our purposes. One is lazy transformations of
collections: an approach semantically equivalent to that of other languages,
which also avoids the creation of intermediate, allocated results. The other
alternative is to use strict collections, which are better supported in
the \scala{} libraries, yet not equivalent to other implementations in our set and
suffering from increased memory consumption.

To achieve lazy processing, one has to use the \sv{view} method on a
collection.\footnote{\scala{} has more APIs for lazy collections
  (e.g., ``Streams''), but the views API we employed is the exact
  counterpart, in spirit and functionality, to the machinery in the
  other languages under study.} This method wraps a collection into a
\sv{SeqView}. The following example illustrates the use of \sv{view}
for performing such transformations lazily:

\noindent\begin{minipage}{\linewidth}\begin{lstlisting}[style=scala]
def sumOfSquareSeq (v : Array[Double]) : Double = {
  val sum : Double = v
  .view
  .map(d => d * d)
  .sum
  sum
}
\end{lstlisting}\end{minipage}

Ultimately, \sv{SeqView} extends \sv{Iterable[A]}, which acts as a factory for
iterators. As an example, we can demonstrate the common \sv{map} function by
mapping the transformation function to the source's \sv{Iterable} iterator:

\noindent\begin{minipage}{\linewidth}\begin{lstlisting}[style=scala]
def map[T, U](source: Iterable[T], f: T => U) = new Iterable[U] {
  def iterator = source.iterator map f
}
\end{lstlisting}\end{minipage} 

The \sv{Iterator}'s \sv{map} function can then be implemented by delegation to
the source iterator:

\noindent\begin{minipage}{\linewidth}\begin{lstlisting}[style=scala]
def map[T, U](source: Iterator[T], f: T => U): Iterator[U] = new Iterator[U] {
  def hasNext = source.hasNext
  def next() = f(source.next())
}
\end{lstlisting}\end{minipage} 

Note that we have 3 virtual calls (\sv{next}, \sv{hasNext}, \sv{f}) per
element pointed by the iterator. The iteration takes place in the
expected, unoptimized order, i.e., each operator has to ``request''
elements from the one supplying its input, rather than having a
``push'' pattern, with the producer calling the consumer directly.

The \scala{} translation is based on synthetic classes that are generated at
compile time. For lambdas, \scala{} generates a class that extends
\sv{scala.runtime.AbstractFunction}. For lambdas with free variables
(captured from the environment), the generated class includes private
member fields that get initialized at instantiation time.

The strict processing of \scala{} collections is similar to the above
lazy idioms from the end-user standpoint: only the \sv{view} call is
omitted in our \sv{sumOfSquareSeq} code example. Operators such as
\sv{map} are overloaded to also process strict collections.

\subsubsection{\cs/\fs}

\cs{} is a modern object-oriented programming language targeting the .NET 
framework. An important milestone for the language was the introduction of
several new major features in \cs{} 3.0 in order to enable a more
functional style of programming. These new features, under the umbrella of
LINQ~\cite{meijer_linq:_2006, meijer_world_2011}, can be summarized as support
for lambda expressions and function closures, extension methods, anonymous types
and special syntax for query comprehensions. All of these new language features
enable the creation of new functional-style APIs for the manipulation of
collections.

\fs{} is a modern .NET functional-first programming language based on
OCaml, with support for object-oriented programming, based on the .NET object system. 

In \cs{} we have two ways of programming with data streams:

1) as fluent-style method calls

\noindent\begin{minipage}{\linewidth}\begin{lstlisting}[style=csharp]
nums.Where(x => x % 2 == 0).Select(x => x * x).Sum();
\end{lstlisting}\end{minipage}

2) or with the equivalent query comprehension syntactic sugar

\noindent\begin{minipage}{\linewidth}\begin{lstlisting}[style=csharp]
(from x in nums
 where x % 2 == 0
 select x * x).Sum();
\end{lstlisting}\end{minipage}

In \fs{}, programming with data is just as simple as a direct pipeline of various
combinators.

%// let (|>) a f = f a 

\noindent\begin{minipage}{\linewidth}\begin{lstlisting}[style=fsharp]
nums  |> Seq.filter (fun x -> x % 2 = 0) 
      |> Seq.map (fun x -> x * x) 
      |> Seq.sum
\end{lstlisting}\end{minipage}

For the purposes of this discussion, we can consider that both \cs{}
and \fs{} have identical operational behaviors and both \cs{} methods
(\sv{Select}, \sv{Where}, etc.) and \fs{} combinators (\sv{Seq.map},
\sv{Seq.filter}, etc.) operate on \sv{IEnumerable<T>} objects and
return \sv{IEnumerable<T>}.

The \sv{IEnumerable<T>} interface can be thought of as a factory for creating
\sv{IEnumerator<T>} objects:

\noindent\begin{minipage}{\linewidth}\begin{lstlisting}[style=csharp]
interface IEnumerable<T> {
  IEnumerator<T> GetEnumerator();
}
\end{lstlisting}\end{minipage}
and \sv{IEnumerator<T>} is an iterator for an on demand consumption of values:

\noindent\begin{minipage}{\linewidth}\begin{lstlisting}[style=csharp]
interface IEnumerator<T> {
  // Return current position element
  T Current { get; }
  // Move to next element,
  // returns false if no more elements remain
  bool MoveNext();
}
\end{lstlisting}\end{minipage}

Each of these methods/combinators implement a pair of interfaces called
\sv{IEnumerable<T>} / \sv{IEnumerator<T>} and through the composition of these
methods a call graph of iterators is chained together. The lazy nature of the
iterators allows the composition of an arbitrary number of operators without
worrying about intermediate materialization of collections between each
call. Instead, each operator call is interleaved with each other. As an example
we can present an implementation of the \sv{Select} method.

\noindent\begin{minipage}{\linewidth}\begin{lstlisting}[style=csharp]
static IEnumerable<R> Select<T, R>(IEnumerable<T> source, Func<T, R> f) {
  return new SelectEnumerable<T, R>(source, f);
}
\end{lstlisting}\end{minipage}

The \sv{SelectEnumerable} has a simple factory-style implementation:

\noindent\begin{minipage}{\linewidth}\begin{lstlisting}[style=csharp]
class SelectEnumerable<T, R> : IEnumerable<R> {	
  private readonly IEnumerable<T> inner;
  private readonly Func<T, R> func;
  public SelectEnumerable(IEnumerable<T> inner, 
                          Func<T, R> func) {
    this.inner = inner;
    this.func = func;
  }
  IEnumerator<R> GetEnumerator() { 
    return new SelectEnumerator(inner.GetEnumerator(), func); 
  }
}
\end{lstlisting}\end{minipage}

\sv{SelectEnumerator} implements the \sv{IEnumerator<R>} interface and delegates
the \sv{MoveNext} and \sv{Current} calls to the inner iterator. 

\noindent\begin{minipage}{\linewidth}\begin{lstlisting}[style=csharp]
class SelectEnumerator<T, R> : IEnumerator<R> {
  private readonly IEnumerator<T> inner;
  private readonly Func<T, R> func;
  public SelectEnumerator(IEnumerator<T> inner, 
                          Func<T, R> func) {
    this.inner = inner;
    this.func = func;
  }
  bool MoveNext() { return inner.MoveNext(); }
  R Current { get { return func(inner.Current); } }
}
\end{lstlisting}\end{minipage}

For programmer convenience, both \cs{} and \fs{} offer support for automatically
creating the elaborate scaffolding of the \sv{IEnumerable<T>} /
\sv{IEnumerator<T>} interfaces, but for our discussion it is not crucial to
understand the mechanisms. 

From a performance point of view, it is not difficult to see that
there is a lot of virtual call indirection between the chained
enumerators. We have 3 virtual calls (\sv{MoveNext}, \sv{Current},
\sv{func}) per element per iterator. Iteration is similar to \scala{}
or to the generic, unoptimized \java{} iteration: it is an external
iteration, with each consumer asking the producer for the next element.

In terms of lambda translation, \cs{} lambdas are always assigned to
delegates, which can be thought of as type-safe function pointers, and
in \fs{} lambdas are represented as compiler generated class types
that inherit \sv{FSharpFunc}.

\noindent\begin{minipage}{\linewidth}\begin{lstlisting}[style=csharp]
abstract class FSharpFunc<T, R> {
  abstract R Invoke(T arg);
}
\end{lstlisting}\end{minipage}

In both cases, if a lambda captures free variables, these variables are
represented as member fields in a compiler-generated class type.

\subsection{Optimizing Frameworks}

We next examine two optimizing frameworks for streaming operations:
ScalaBlitz and LinqOptimizer.

\subsubsection{ScalaBlitz}

ScalaBlitz is an open source framework that optimizes Scala
collections by applying optimizations for both sequential and parallel
computations. It eliminates boxing, performs lambda inlining, loop
fusion and specializations to particular data-structures. ScalaBlitz
performs optimizations at compile-time based on Scala
macros~\cite{burmako_scala_2013}.

By enclosing a functional pipeline into an \sv{optimize} block, ScalaBlitz
expands in place an optimized version of it:

\noindent\begin{minipage}{\linewidth}\begin{lstlisting}[style=scala]
def sumOfSquareSeqBlitz (v : Array[Double]) : Double = {
  optimize {
    val sum : Double = v
    .map(d => d * d)
    .sum
    sum
  }
}
\end{lstlisting}\end{minipage}

This can be achieved because this library is implemented as a def macro with the
following signature:

\noindent\begin{minipage}{\linewidth}\begin{lstlisting}[style=scala]
def optimize[T](exp: T): Any = macro optimize_impl[T]
\end{lstlisting}\end{minipage}

The \sv{optimize} block is a function that starts with the additional keyword
\sv{macro}. When the compiler encounters an application of the macro
\sv{optimize(expression)}, it will expand that application by invoking
\sv{optimize\_impl}, with the abstract-syntax tree of the functional pipeline
expression as argument. The result of the macro implementation is an expanded
abstract syntax tree. This tree will be replaced at the call site and will be
type-checked.

\subsubsection{LinqOptimizer}
LinqOptimizer is an open source optimizer for LINQ queries.
It compiles declarative queries into fast loop-based imperative code,
eliminating virtual calls and temporary heap
allocations. LinqOptimizer is a run-time compiler based on
LINQ Expression trees, which enable a form of metaprogramming based on
type-directed quotations. 

In the following example, a lambda expression is assigned to a
variable of type \sv{Expression<Func<\ldots>>}. 

\noindent
\begin{minipage}{\linewidth}\begin{lstlisting}[style=csharp]
Expression<Func<int, int>> exprf = x => x + 1;
Func<int, int> f = exprf.Compile(); // compile to IL
Console.WriteLine(f(1)); // 2
\end{lstlisting}\end{minipage}

At compile time, the
compiler emits code to build an expression tree that represents the
lambda expression. LINQ offers library support for runtime
manipulation of expression trees (through visitors) and also support
for run-time compilation to IL.
Using such features, LinqOptimizer lifts queries into the world of
expression trees and performs the following optimizations:

1) inlines lambdas and performs loop fusion:

\begin{lstlisting}[style=csharp]
var sum = (from num in nums.AsQueryExpr() // lift
           where num % 2 == 0
           select num * num).Sum();
// effectively optimizes to
int sum = 0;
for (int index = 0; index < nums.Length; index++) {
   int num = nums[index];
   if (num % 2 == 0)
      sum += num * num;
}
\end{lstlisting}

2) for queries with nested structure (\sv{SelectMany}, \sv{flatMap}) applies
nested loop generation:

\noindent
\begin{minipage}{\linewidth}\begin{lstlisting}[style=csharp]
var sum = (from num in nums.AsQueryExpr() // lift
           from _num in _nums
           where num % 2 == 0
           select num * _num).Sum();
// effectively optimizes to
int sum = 0;
for (int index = 0; index < nums.Length; index++) {
   for (int _index = 0; _index < _nums.Length; 
        _index++) {
      int num = nums[index];
      int _num = _nums[_index];
      if (num % 2 == 0)
         sum += num * _num;
   }
}
\end{lstlisting}\end{minipage}

\section{Results}

We next discuss our benchmarks and experimental results.

\subsection{Microbenchmarks}
In this work, we use 4 main microbenchmarks. We focus our efforts on
measuring iteration throughput and lambda invocation costs. In all of
our benchmarks we produce scalar values as the result of a terminal
operation (e.g., instead of producing a transformed list of values), as
we do not want to cause memory management effects (e.g., garbage collection).
Furthermore, we did not employ sorting or grouping operators, in
order to avoid interfering with algorithmic details of library
implementations (e.g., mergesort vs quicksort, hash tables vs balanced
trees, etc.). 

We measure the performance of:

\begin{itemize}
 \item \textbf{sum} iteration speed with no lambdas, just a single iteration.
 \item \textbf{sumOfSquares} a small pipeline with one map operation (i.e., one
   lambda).
 \item \textbf{sumOfSquaresEven} a bigger pipeline with a filter and map chain of
   two lambdas.
 \item \textbf{cart} a nested pipeline with a \sv{flatMap} and an inner operation,
   again with a \sv{flatMap} (capturing a variable), to encode a Cartesian product.
\end{itemize}

We developed this set for all four languages, \java{}, \scala{}, \cs{} and
\fs{}, for both sequential and parallel execution. For the latter three we have
also included optimized versions using ScalaBlitz and LinqOptimizer. For
\scala{} we also include alternate implementations, which employ more idiomatic
strict collections (without the views API). Arguably this approach is better
supported in the \scala{} libraries. Therefore we present separate measurements
for \scala{}-views and \scala{}-strict tests. In our following analysis, when we do not refer to a
\scala{}-strict test explicitly, \scala{}-views are implied. Additionally, we
include a baseline suite of benchmarks for the sequential cases.

We have run these benchmarks on both Windows and Linux, although
Windows is the more universal reference platform for our comparison:
it allows us to perform the \cs{}/\fs{} tests on the
industrial-strength implementation of the Microsoft CLR virtual
machine.

The purpose of baseline benchmarks is to assess the performance difference
between functional pipelines and the corresponding imperative, hand-written
equivalents. The imperative examples make use of indexed-based loop iterations
in the form of \sv{for}-loops (except for the \scala{} case in which the while-loop
is the analogue of imperative iteration). 

\paragraph{Input:} All tests were run with the same input set. For the
\textbf{sum}, \textbf{sumOfSquares} and \textbf{sumOfSquaresEven} we used an
array of $N = 10,000,000$ long integers, produced by $N$ integers with a \sv{range}
function. The \textbf{cart} test iterates over two arrays. An outer
one of $1,000,000$ long integers and an inner one of $10$.

The \scala{}, \cs{} and \fs{} tests were compiled with optimization
flags enabled and for \java{}/\scala{} tiered compilation was left
disabled (C2 JIT compiler only). Additionally, we fixed the heap size
to 3GB for the JVM to avoid heap resizing effects during
execution.

\subsection{Experimental Setup} 
\begin{center}
\begin{tabular}{ c c c}
  \toprule
  & Windows & Ubuntu Linux \\
  \midrule
  Version & 8.1 & 13.10/3.11.0-24 \\ 
  Architecture & x64 & x64 \\
  CPU & \multicolumn{2}{c}{Intel Core i5-3360M vPro 2.8GHz} \\
  Cores & \multicolumn{2}{c}{2 physical x 2 logical} \\
  Memory & \multicolumn{2}{c}{4GB} \\
  \bottomrule
\end{tabular}
\end{center}

\begin{figure*}
  \center
  \includegraphics[width=10cm]{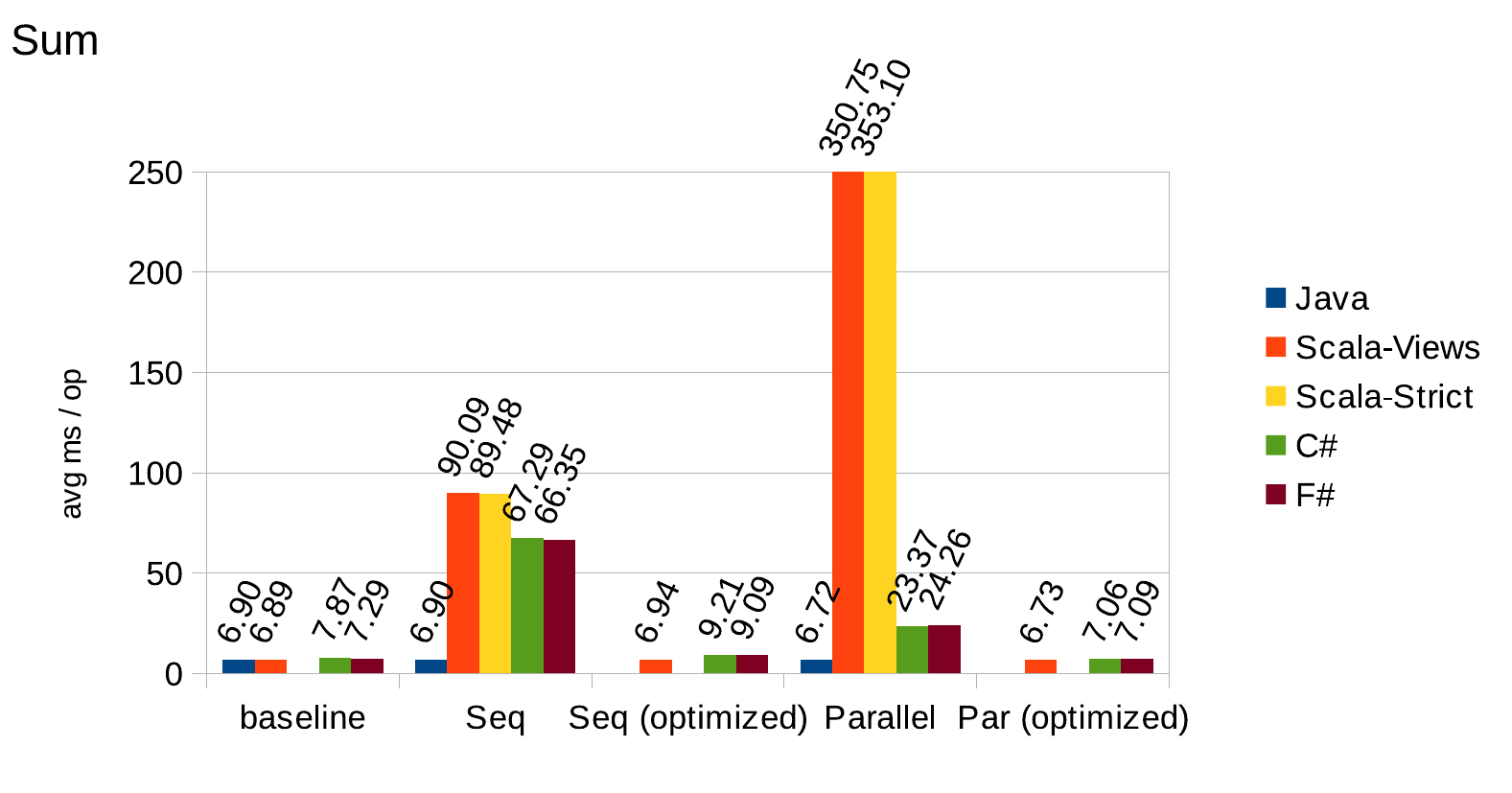}
  \includegraphics[width=10cm]{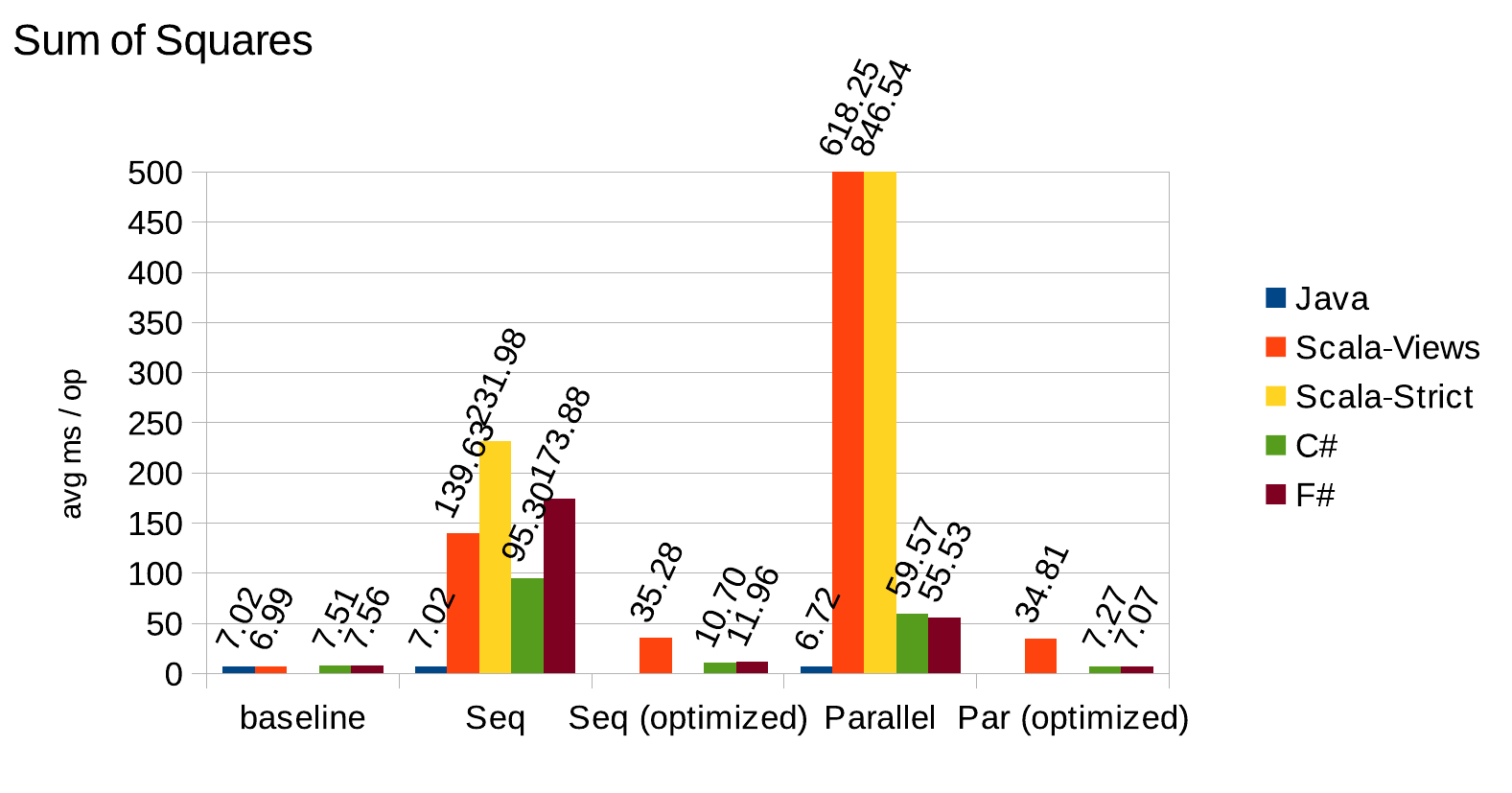}
  \includegraphics[width=10cm]{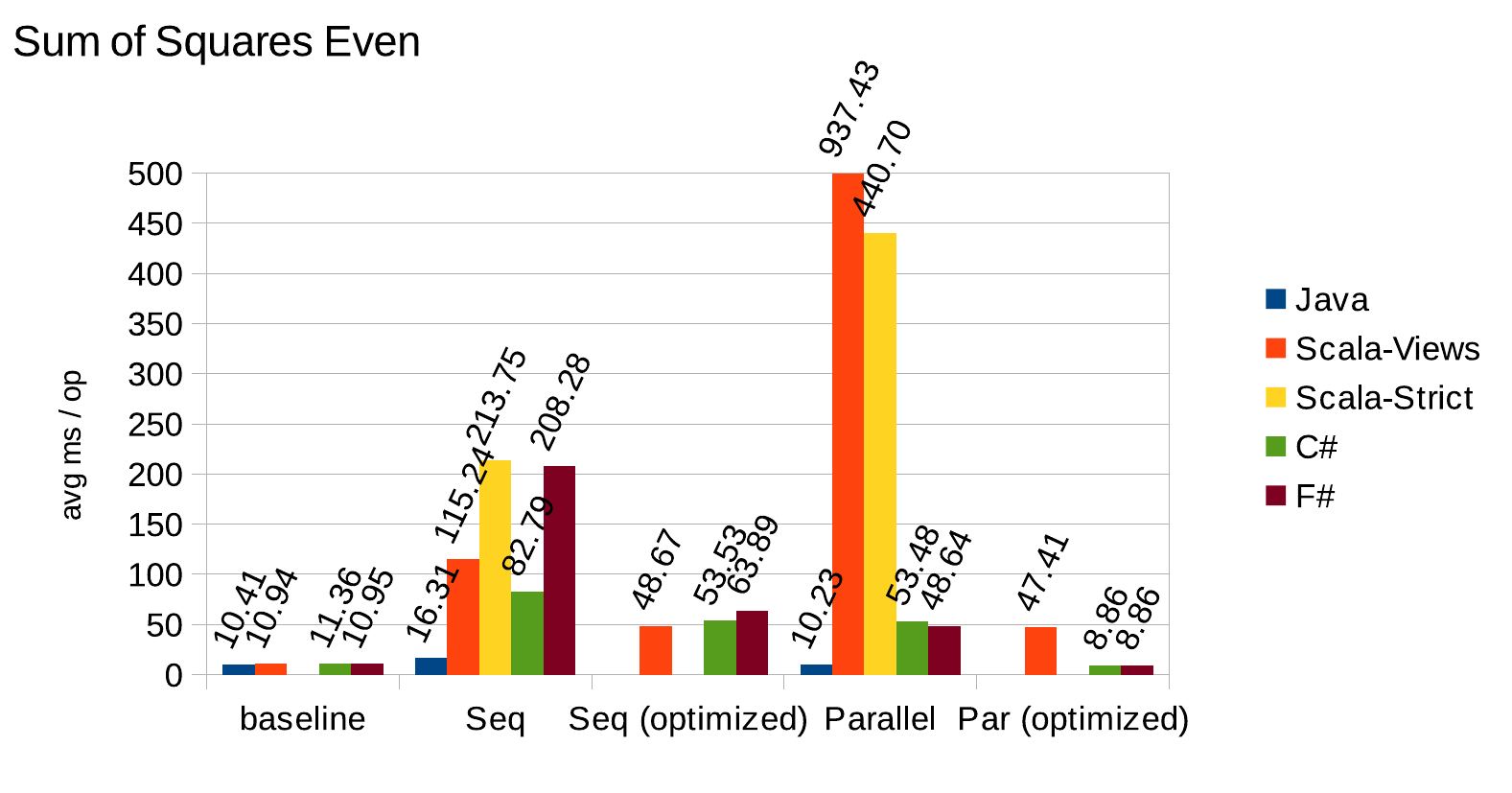}
  \includegraphics[width=10cm]{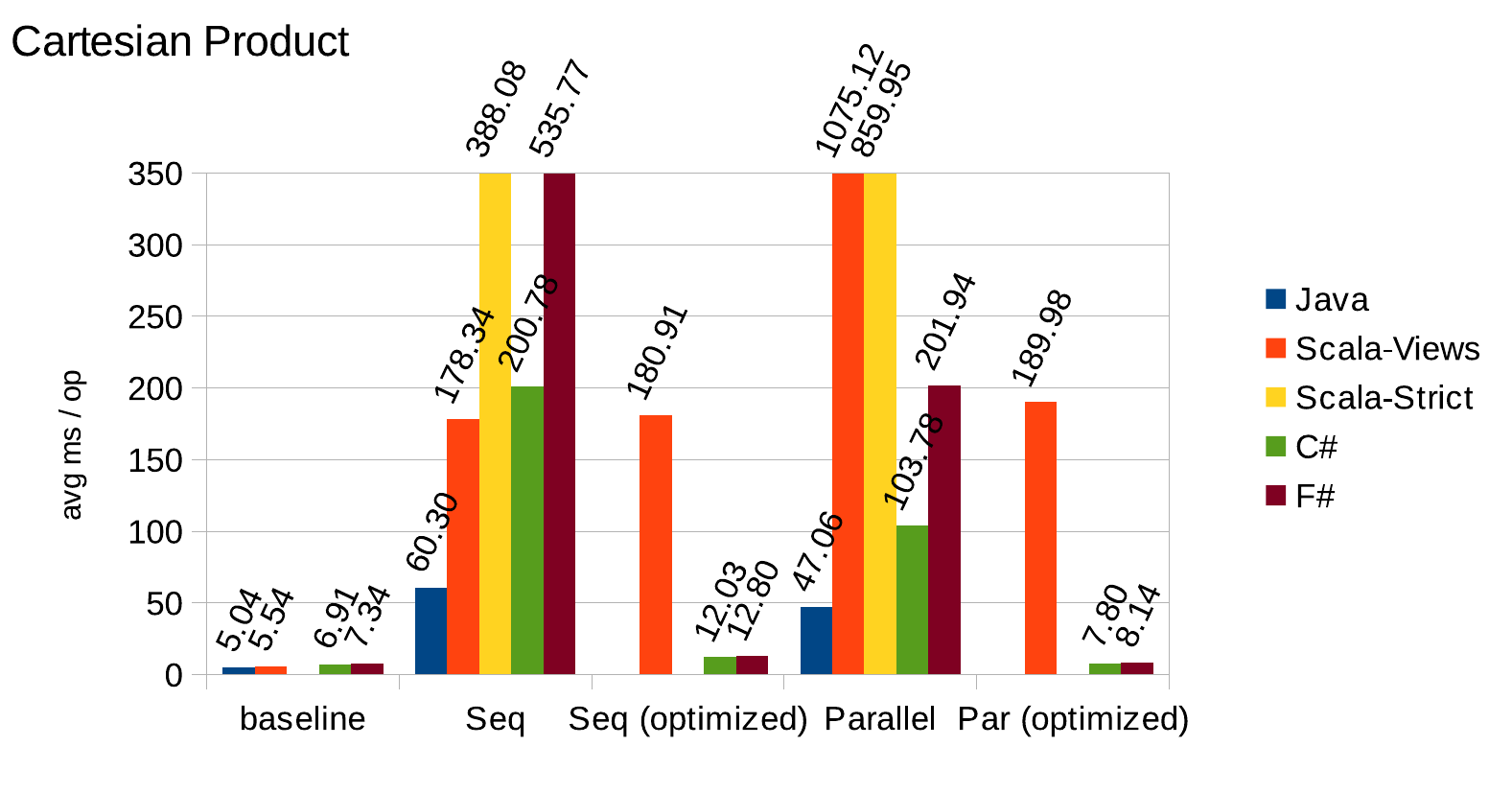}
  \caption{Microbenchmark Results on Windows (CLR/JVM) in milliseconds / iteration (average of 10). Y-axis truncated for readability.}
  \label{fig:windows}
\end{figure*}

\begin{figure*}
  \center
  \includegraphics[width=10cm]{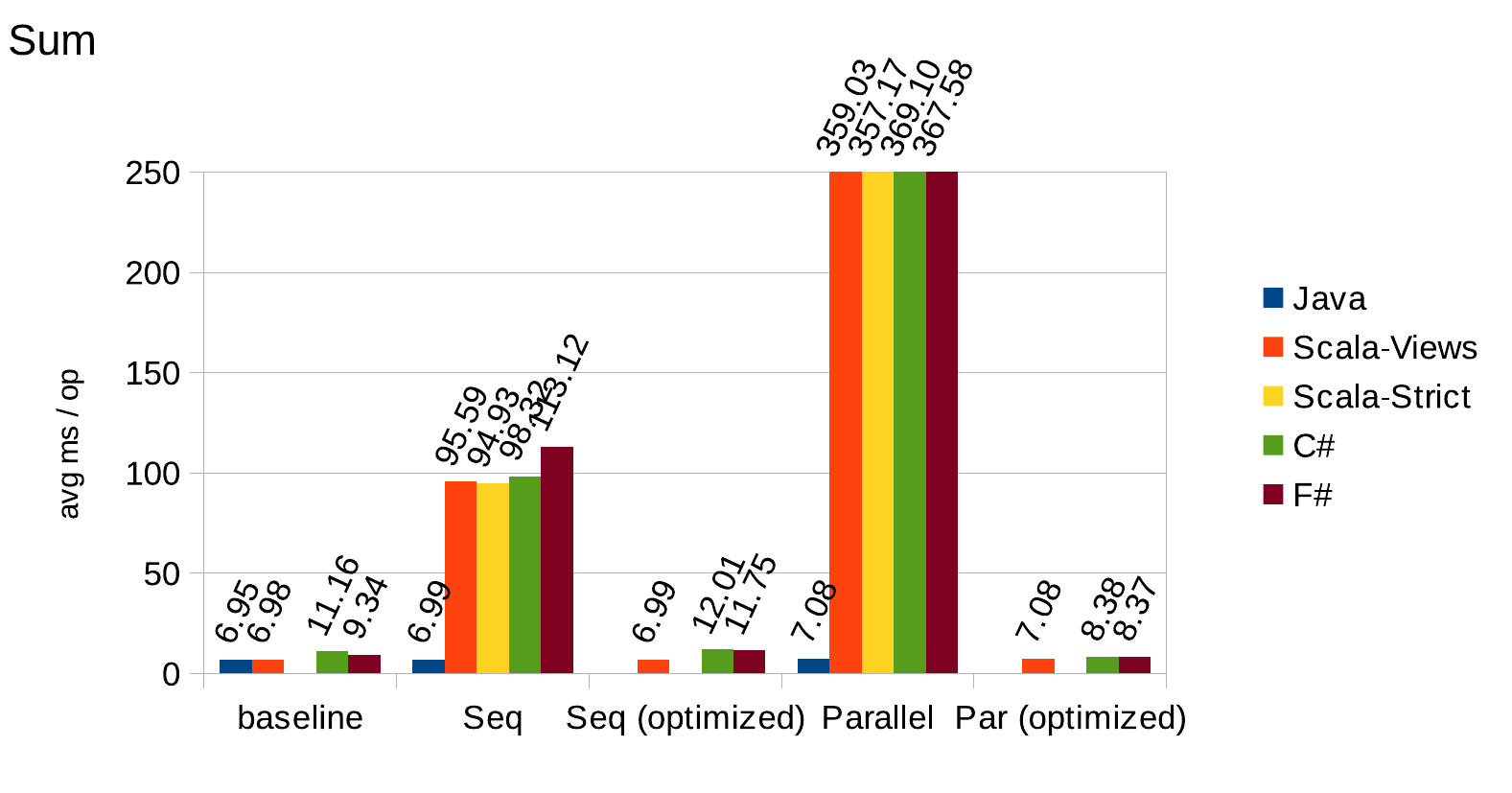}
  \includegraphics[width=10cm]{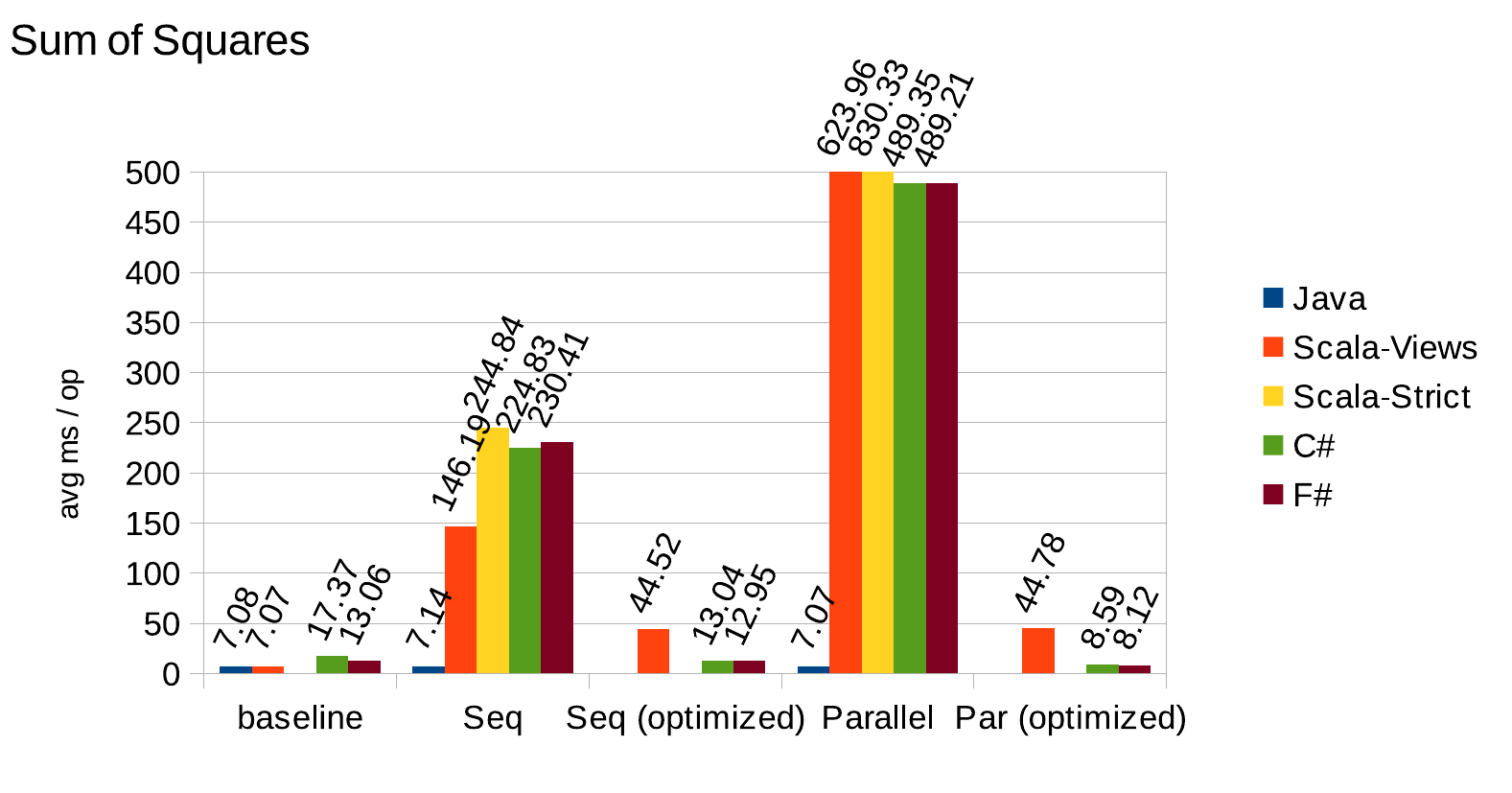}
  \includegraphics[width=10cm]{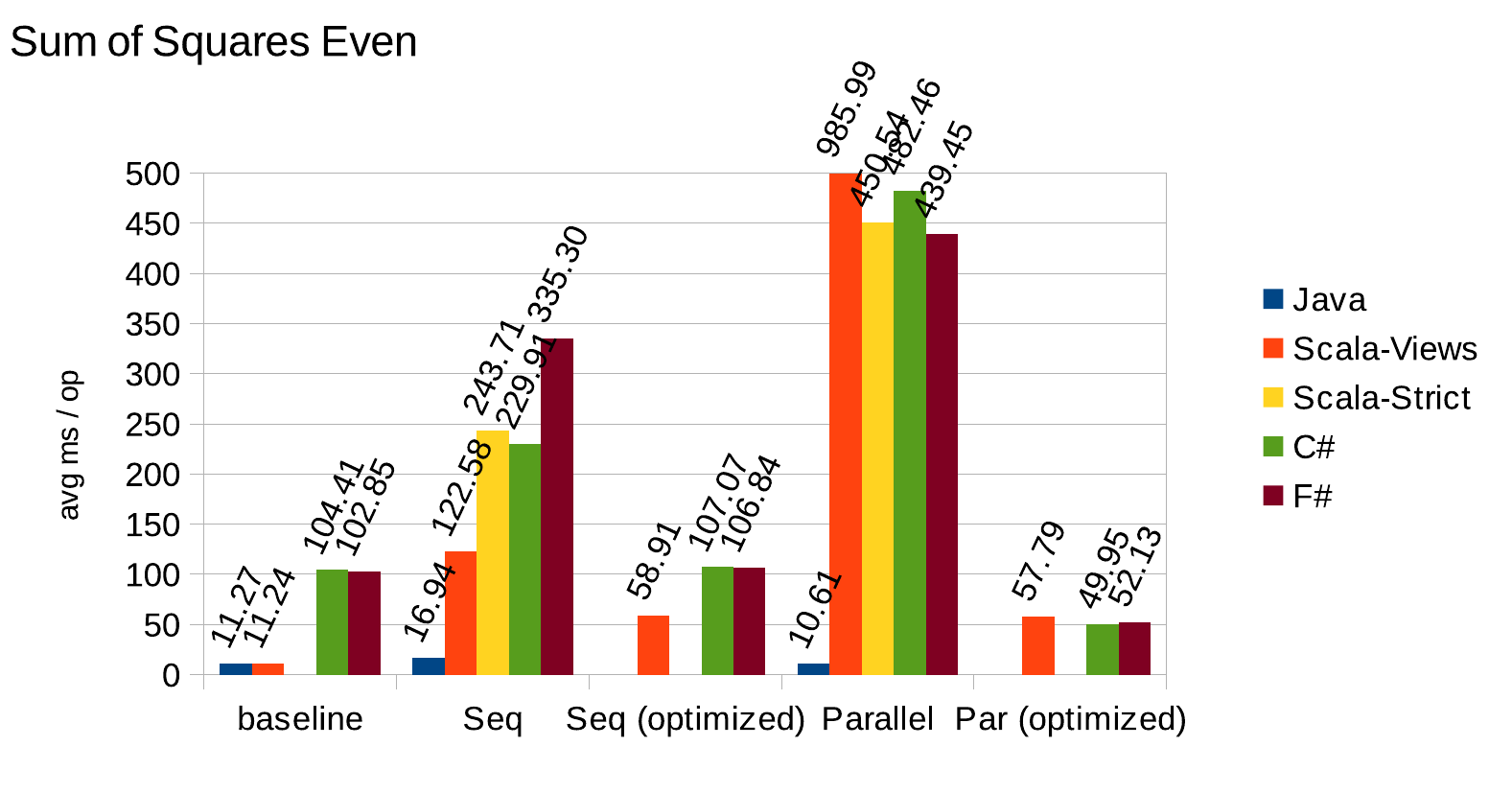}
  \includegraphics[width=10cm]{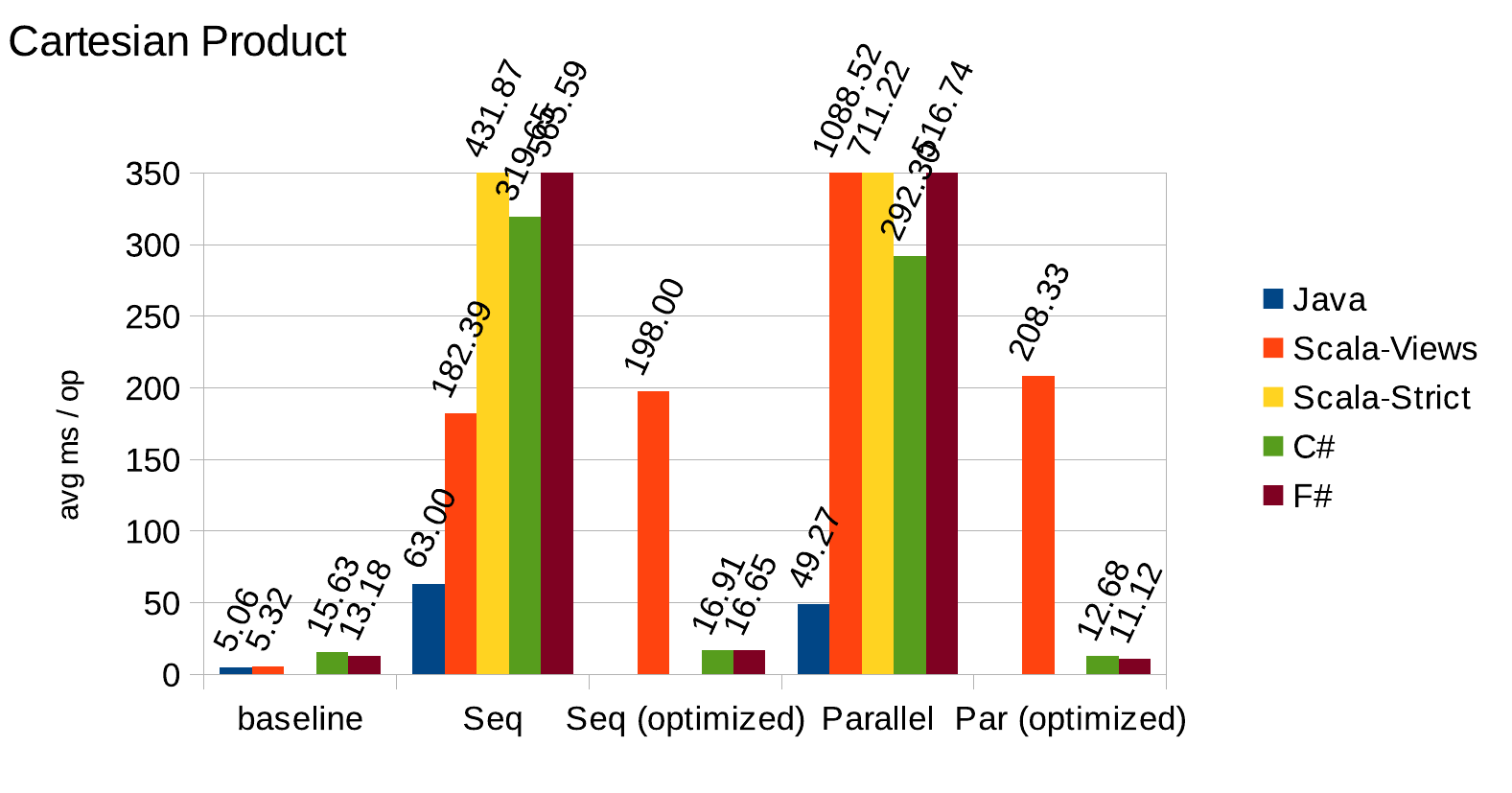}
  \caption{Microbenchmark Results on Linux (mono/JVM) in milliseconds / iteration (average of 10). Y-axis truncated for readability.}
  \label{fig:linux}
\end{figure*}

\paragraph{Systems:} We performed 
both Linux (see Figure~\ref{fig:linux}) and Windows (see
Figure~\ref{fig:windows}) tests natively on the same system via a dual-boot
installation. 

\begin{center}
\begin{tabular}{ c c c }
  \toprule
  & Windows & Ubuntu Linux \\
  \midrule
  Java & \multicolumn{2}{c}{Java 8 (b132)/JVM 1.8} \\ 
  Scala & \multicolumn{2}{c}{2.10.4/JVM 1.8} \\
  \cs{} & \cs{}5  /CLR v4.0 & \cs{} mono 3.4.0.0/mono 3.4.0 \\
  \fs{} & \fs{}3.1/CLR v4.0 & \fs{} open-source 3.0/mono 3.4.0 \\
  \bottomrule
\end{tabular}
\end{center}

\paragraph{Microbenchmarking automation:} 
For Java and \scala{} benchmarks we used the Java Microbenchmark Harness
(JMH)~\cite{aleksey_shipilev_openjdk:_????} tool: a benchmarking tool for
JVM-based languages that is part of the OpenJDK. JMH is an annotation-based tool
and takes care of all intrinsic details of the execution process. Its goal is to
produce as objective results as possible. The JVM performs JIT compilation (we use the C2 JIT compiler)
so the benchmark author must
measure execution time after a certain warm-up period to wait for transient
responses to settle down. JMH offers an easy API to achieve that. In our
benchmarks we employed 10 warm-up iterations and 10 proper iterations. We also
force garbage collection before benchmark execution. Regarding the CLR, warm-up
effects take an infinitesimal amount of time compared to the
JVM~\cite{singer_jvm_2003}. The CLR JIT compiler compiles methods exactly once
and subsequent method calls invoke directly the JITted version. Code is never
recompiled (nor interpreted at any point). For the purpose of benchmarking
\cs{}/\fs{} programs, as there is not any widely-used, state-of-the-art tool for
microbenchmarking, we created the~\emph{LambdaMicrobenchmarking}~utility~\footnote{https://github.com/biboudis/LambdaMicrobenchmarking .}
written in \cs{}, according to the common microbenchmarking practices described
in~\cite{sestoft2013microbenchmarks}. It calculates the average execution of method invocations
using the \sv{TimeSpan.TotalMilliseconds} property of the \sv{TimeSpan}
structure that converts ticks to whole and fractional milliseconds. Our utility
uses the Student-T distribution for statistical inference; mean error and
standard deviation. The same distribution is employed in JMH as well. Our
utility forces garbage collection between runs. For all tests, we do not measure
the time needed to initialize data-structures (filling arrays), 
and neither the run-time compilation cost of the optimized queries in the
LinqOptimizer case nor the compile-time overhead of macro expansion in the
ScalaBlitz case.

\subsection{Performance Evaluation}

\paragraph{Languages:}
Among the languages\footnote{Although it is easy to categorize
  benchmarks per language, and we refer to languages throughout, it is
  important to keep in mind that the comparison concerns primarily the
  standard libraries of these languages and only secondarily the
  language translation techniques for lambdas.} of our study, \java{}
exhibits by far the best performance, in both sequential and parallel
tests, due to its advanced translation scheme. Notably, \java{}
results show not only that three out of four of our tests are very
close to baseline measurements but also that the parallel versions
scale well. Regarding the parallel versions, all microbenchmarks
reveal that even in cases where \java{} was very close to the
baseline, performance increases further achieving parallel speedups of
1.1x-1.6x.  For the \textbf{cart} benchmark, although Java has the best
performance among all streaming implementations, it still pays a
considerable cost for inner closures, as can be seen in comparison to
the baseline benchmark for the sequential case. During the execution
of \textbf{cart} the garbage collector was
invoked 3 times (per iteration) for the sequential version and 4 times for the parallel version,
indicating significant memory management activity.
\newcommand{\subhead}[1]{\multicolumn{1}{c}{#1}}% to format sub-headings of d-type columns
\newcommand{\ra}[1]{\renewcommand{\arraystretch}{#1}}
\newcolumntype{d}[1]{D{.}{.}{4}}
\begin{table*}\centering
  {\scriptsize
    \begin{tabular}{ l d{4} d{4} d{4} d{4} d{4} d{4} d{4} d{4} d{4} d{4} }
      \toprule
      & \multicolumn{5}{c}{\textbf{Windows}} & \multicolumn{5}{c}{\textbf{Linux}} \\
      \cmidrule(rl){2-6} \cmidrule(rl){7-11}
      Benchmark & \subhead{Java} & \subhead{Scala-Views} & \subhead{Scala-Strict} &  \subhead{C\#} &  \subhead{F\#} & \subhead{Java} & \subhead{Scala-Views} & \subhead{Scala-Strict} & \subhead{C\#} &  \subhead{F\#} \\ 
      \midrule
      sumBaseline & 0.011 & 0.015 &  & 1.214 & 0.168 & 0.054 & 0.011 &  & 0.552 & 0.818\\
      sumSeq & 0.015 & 0.607 & 0.277 & 2.407 & 0.525 & 0.014 & 0.449 & 0.475 & 0.359 & 1.015\\
      sumSeqOpt &  & 0.010 &  & 0.536 & 0.212 &  & 0.022 &  & 0.248 & 0.730\\
      sumPar & 0.035 & 2.348 & 2.622 & 0.895 & 4.371 & 0.009 & 3.653 & 1.827 & 106.800 & 117.358\\
      sumParOpt &  & 0.017 &  & 0.075 & 0.196 &  & 0.026 &  & 1.400 & 2.010\\
      sumOfSquaresBaseline & 0.008 & 0.016 &  & 0.129 & 0.202 & 0.023 & 0.013 &  & 0.799 & 1.072\\
      sumOfSquaresSeq & 0.009 & 1.049 & 2.052 & 0.763 & 3.755 & 0.019 & 1.331 & 0.895 & 1.193 & 1.116\\
      sumOfSquaresSeqOpt &  & 1.104 &  & 0.215 & 0.292 &  & 0.238 &  & 0.583 & 0.171\\
      sumOfSquaresPar & 0.008 & 3.691 & 9.355 & 2.745 & 0.162 & 0.017 & 2.807 & 6.347 & 23.856 & 40.342\\
      sumOfSquaresParOpt &  & 0.036 &  & 0.433 & 0.094 &  & 0.136 &  & 0.782 & 0.485\\
      sumOfSquaresEvenBaseline & 0.044 & 0.085 &  & 0.204 & 0.393 & 0.059 & 0.035 &  & 0.906 & 1.270\\
      sumOfSquaresEvenSeq & 0.121 & 1.157 & 1.510 & 3.789 & 4.838 & 0.096 & 1.159 & 1.042 & 0.895 & 1.680\\
      sumOfSquaresEvenSeqOpt &  & 0.550 &  & 2.052 & 5.351 &  & 0.162 &  & 0.847 & 0.522\\
      sumOfSquaresEvenPar & 0.025 & 5.184 & 8.207 & 5.943 & 2.556 & 0.027 & 4.905 & 16.252 & 46.739 & 21.465\\
      sumOfSquaresEvenParOpt &  & 0.502 &  & 0.115 & 0.128 &  & 0.483 &  & 1.737 & 4.390\\
      cartBaseline & 0.060 & 0.041 &  & 0.015 & 1.007 & 0.010 & 0.010 &  & 0.040 & 0.113\\
      cartSeq & 0.749 & 6.195 & 3.939 & 4.284 & 5.840 & 0.510 & 2.437 & 5.486 & 0.954 & 2.791\\
      cartSeqOpt &  & 0.666 &  & 0.148 & 0.232 &  & 0.763 &  & 0.751 & 0.307\\
      cartPar & 0.131 & 13.167 & 13.165 & 4.954 & 7.855 & 0.243 & 7.641 & 7.484 & 10.963 & 7.546\\
      cartParOpt &  & 2.694 &  & 0.904 & 1.371 &  & 2.642 &  & 1.810 & 1.310\\
      refBaseline & 0.069 & 0.259 &  & 0.159 & 0.360 & 0.152 & 0.288 &  & 1.740 & 1.566\\
      refSeq & 0.221 & 1.077 & 0.719 & 1.267 & 3.415 & 0.237 & 0.438 & 0.353 & 1.269 & 0.639\\
      refSeqOpt &  & 0.284 &  & 2.082 & 1.437 &  & 0.235 &  & 2.409 & 1.643\\
      refPar & 0.119 & 5.123 & 0.853 & 8.548 & 2.556 & 0.271 & 6.904 & 0.765 & 44.879 & 27.644\\
      refParOpt &  & 0.247 &  & 0.782 & 0.187 &  & 0.112 &  & 1.445 & 2.592\\
      \bottomrule
    \end{tabular}
}
\caption{Standard deviations for 10 runs of each benchmark.}
\label{sdevs}
\end{table*}
\scala{} Parallel Collections using the lazy, \sv{view}, API seem to
suffer in the parallel tests quite significantly over all other
implementations (note that the Y-axis is truncated) due to
boxing/unboxing, iterator, and function object abstraction
penalties. (For a more detailed analysis, see Section~\ref{sec:discussion}.)
The strict \scala{} API (which, although
non-equivalent to other implementations is arguably more idiomatic)
performed significantly better. Although we present results
for a 3GB heap space, we have also conducted the same tests
under various constrained heap spaces. In practice,
\scala{}-strict benchmarks ran with about 4x more heap space than
their \java{} counterparts, which is unsurprising given that all
strict operators need to generate and process intermediate
collections. Still, the parallel \scala{}/\scala{}-strict benchmarks
were almost always the slowest among all implementations on both Windows
and Linux.% (with the exception of one \cs{} and \fs{} benchmark on mono).

In the sequential tests of \cs{} and \fs{} we observe a constant difference in
favor of \cs{} for \textbf{sumOfSquares}, \textbf{sumOfSquaresEven} and a
significant difference of 2.7x for the \textbf{cart} benchmark on Windows.
As \sv{seq<\textquotesingle T>} is a type alias for .NET's \sv{IEnumerable<T>} we
conclude that the difference is attributed to different
implementations of operators. In the parallel benchmarks, as \fs{} relies on the standard
library for .NET, it is driven by its performance. Thus, all
parallel benchmarks (Windows and Linux) show these two languages at the same
level.

In all cases, the parallel benchmarks of LINQ on mono scaled poorly, revealing
poor scaling decisions in the implementation. Additionally,
comparing the Windows and Linux charts for the respective baseline benchmarks, mono
seems to have generated slower code for the \textbf{sumOfSquaresEven} benchmark, in
which the modulo operation is applied. This indicates that JIT compilation
optimizations can be improved significantly, especially in cases such as the
handwritten fused loop-if operation of the \textbf{sumOfSquaresEven} situation.

Among all standard parallel libraries, \fs{} achieved the best scaling of
2.6x-4.3x.

\paragraph{Optimizing frameworks:}
When streaming pipelines are amenable to optimization, the improvement can be
dramatic. 

ScalaBlitz improved \scala{} in virtually all cases. Especially in the \textbf{sum} benchmark, \scala{} was 
significantly improved, achieving an execution time close to that of the
\java{}/\scala{} baseline tests. Notable are the 52x speed-up in relation to \scala{}
Parallel Collections for the \textbf{sum} benchmark on Windows, as well as 50x on
Linux. Additionally, ScalaBlitz achieved a 17x improvement for
\textbf{sumOfSquares} and 19x for \textbf{sumOfSquaresEven} (again for the parallel
benchmarks) on Windows. ScalaBlitz did not demonstrate improved performance in the
case of nested loops (sequential \textbf{cart}) but presented a 5.7x speedup in
the parallel version on Windows (and 5.2x on Linux). Apart from the elimination of
abstraction penalties, ScalaBlitz offers additional performance improvement in 
the parallel optimized versions
% (in relation to standard ones) 
due to its iterators that allow fine-grained and efficient
work-stealing~\cite{prokopec_lock-free_2014}.

LinqOptimizer improved in all cases the performance of the \cs{} and
\fs{} benchmarks. The result of LinqOptimizer universally demonstrates
the smallest performance gap with the baseline benchmarks, in absolute
values. Especially in the \textbf{cart} benchmark, LinqOptimizer achieved a speed-up
of 17x(sequential) and 13x(parallel) for \cs{} and 42x and 25x respectively for \fs{}. Among the two .NET languages,
\fs{} is the one that benefits more by LinqOptimizer in the sequential
\textbf{sumOfSquares} and \textbf{sumOfSquaresEven} benchmarks. \fs{} gets 14x and 3x
improvements for these benchmarks, respectively, while \cs{} gets 9x
and 1.5x for the sequential tests. In the case of \textbf{cart}, LinqOptimizer has employed the nested
loop optimization, which brings execution near the baseline level.

In table~\ref{sdevs} we present the standard deviation of all
microbenchmarks. Among all measurements, the parallel collections of
\scala{} and \cs{}/ \fs{} on mono/Linux presented the highest
deviations. \java{} demonstrates the highest stability. The strict
version of \scala{} for the parallel \textbf{sumOfSquares} benchmark
exhibit a relatively higher standard deviation, possibly because of
memory effects.

\section{Discussion}
\label{sec:discussion}

Our microbenchmarks paint a fairly clear picture of the current status
of lambdas+streaming implementations, as well as their future improvement
prospects. \java{} employs the most aggressive implementation
technique that does not perform invasive optimization. Other
languages could benefit from the same translation approach. At the
same time, \java{} does not have an optimization framework along the
lines of ScalaBlitz or LinqOptimizer. The \textbf{cart} microbenchmark
showcases the need for such optimizations: \cs{}/\fs{} are 7x faster
in parallel performance than \java{}. For more realistic programs,
such benefits may arise more often. Hence, identifying cases in which
\java{} can benefit from a Stream API optimizing framework (as in the
closed-over variables of \textbf{cart}) is a promising direction. 
\begin{figure*}
\centering
\begin{subfigure}{.5\textwidth}
  \centering
  \includegraphics[width=.95\linewidth]{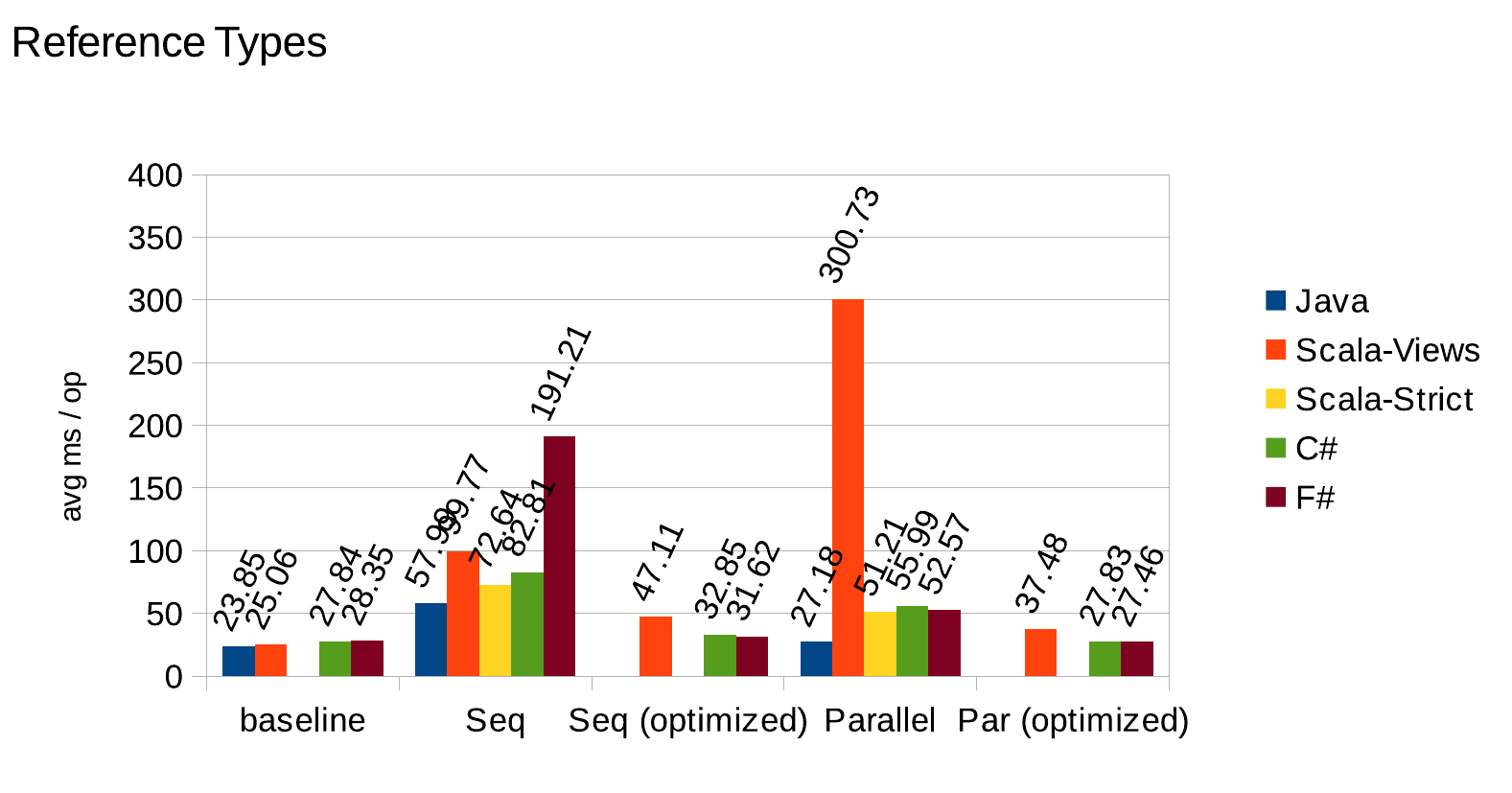}
  \caption{Windows tests}
  \label{fig:sub1}
\end{subfigure}%
\begin{subfigure}{.5\textwidth}
  \centering
  \includegraphics[width=.95\linewidth]{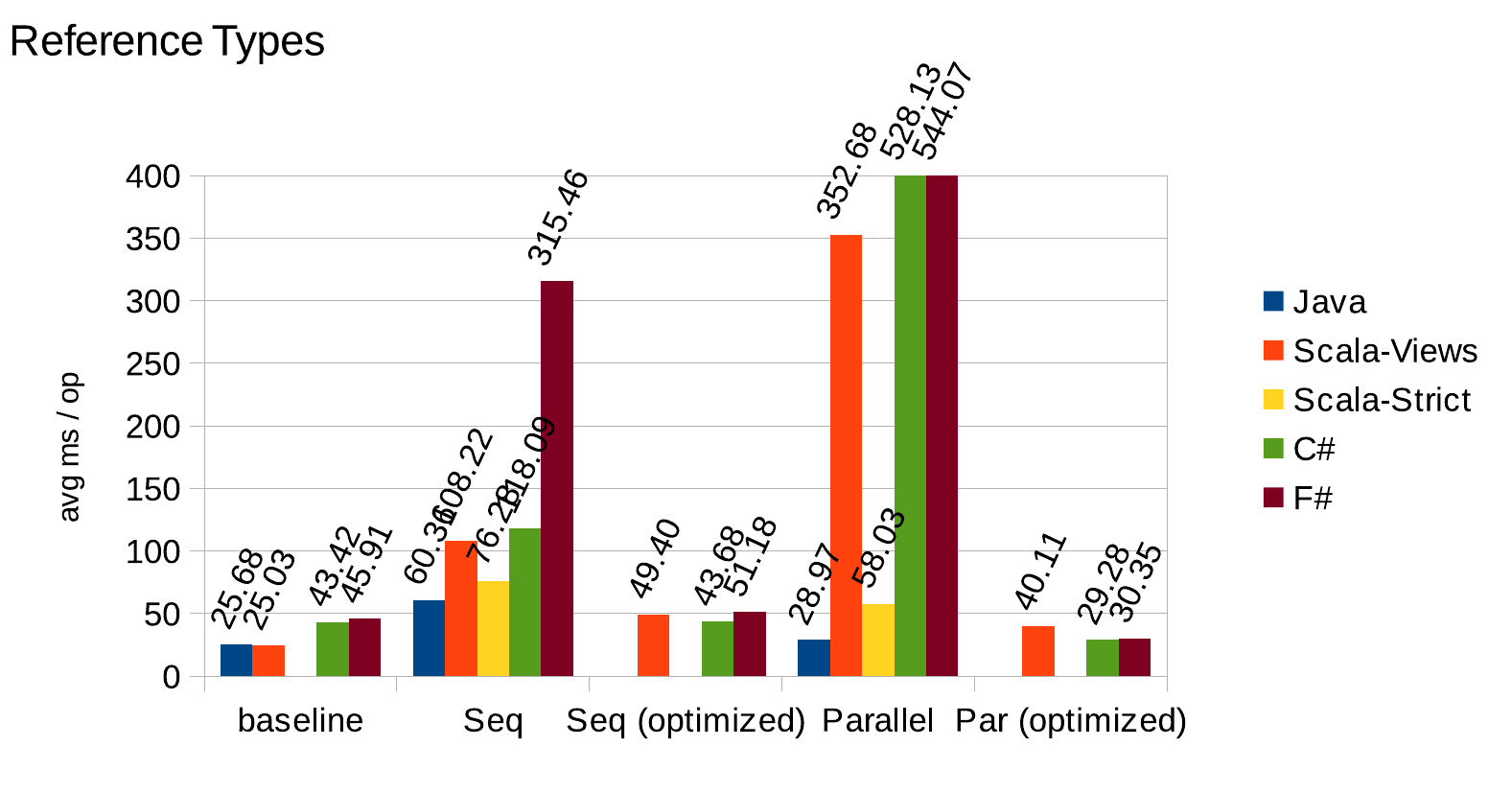}
  \caption{Linux tests}
  \label{fig:sub2}
\end{subfigure}
\caption{Microbenchmark with manual boxing. Y-axis in milliseconds / iteration (average of 10), truncated for readability.}
\label{refs}
\end{figure*}
\scala{} is an outlier in most of our measurements. We found that its
performance, in both the strict and the non-strict case, is subject to
memory management effects. We first examined whether such effects can
be alleviated with the use of VM flags, without intrusive changes to
the benchmarks' source code.  Our microbenchmark runs employ the
default JVM setup of a parallel garbage collector (GC) with \emph{GC
  ergonomics} enabled by default. GC ergonomics is an adaptive
mechanism that tries to meet (in order) three goals: 1) minimize pause
time, 2) maximize throughput, 3) minimize footprint. Leaving GC
ergonomics enabled is not always beneficial for \scala{}.  We
conducted the same tests without the use of adaptive sizing
(-XX:-UseAdaptiveSizePolicy) and no explicit sizing of generations (on
Linux). For both strict and non-strict (not optimized) parallel tests,
we observed an improvement of 1.1x-2.9x, with the parallel version of
\textbf{sumOfSquaresPar} exhibiting the maximum increase. However,
removing adaptive sizing of the heap also causes a performance
degradation of about 10\%-15\% in the majority of sequential tests.
In limited exploration (also based on suggestions by \scala{}
experts) we found no other flag setup that significantly affects
performance.

The main problem with \scala{} performance is that the \scala{}
Collections are not specialized for primitive types. Therefore, \scala{} suffers
significant boxing and unboxing overheads for primitive values, as
well as memory pressure due to the creation of intermediate (boxed)
objects.  Prokopec et al.~\cite{prokopec_lock-free_2014} explain such
issues, along with the effects of indirections and iterator
performance.  Method-level specialization for primitive types can
currently be effected in two ways. One is the \scala{}
\sv{@specialized} annotation, which specializes chains of annotated
generic call sites~\cite{4820/THESES}, while the other is
\emph{Miniboxing}~\cite{miniboxing}. Use of the \sv{@specialized}
annotation causes the injection of specialized method calls while
preserving compatibility with generic code. The use of
\sv{@specialized} preserves separate compilation by generating all
variants of specialized methods, hence leading to bytecode
explosion. Partly due to such considerations, \scala{} Collections do
not employ the \sv{@specialized} annotation. Miniboxing presents a
promising alternative that minimizes bytecode size and defers
transformations to load time. Currently Miniboxing is offered as a
\scala{} compiler plugin. Having specialized collections in the \scala{} 
standard library could greatly improve performance in our benchmarks. 

To demonstrate the above points, in Figure~\ref{refs} we present an
additional benchmark (\textbf{refs}), which executes a pipeline with
reference types and avoids automatic boxing of our input data. The
benchmark operates on an array of $10,000,000$ instances of a class,
\sv{Ref}, employs two filter combinators, and finally returns the size
of the resulting collection. This benchmark effectively performs
boxing manually, for all languages. In this benchmark, \java{}
outperforms other streaming libraries but the difference is quite
small. \scala{} is now directly comparable to all other
implementations, since it performs no extraneous boxing compared to
other languages. Both sequential and parallel tests for \java{} didn't
invoke the GC. However, \scala{} in the \sv{Filtered} trait, which is
defined in the \sv{GenSeqViewLike} implementation trait, causes
internal boxing for the size operator. The \sv{length} definition in
\sv{Filtered}, which delegates to the lazy value of \sv{index}, and
the array allocation inside that lazy value are responsible for this
effect. In the \scala{}-strict parallel test, nearly 100\% of the
allocated memory (originating both from the main thread and from the
Fork/Join workers) comes from the intermediate arrays, but the
ample heap space combined with the almost perfect inlining of the
main internal transformer
(\sv{ParArrayIterator.filter2combiner\_quick}) makes the \scala{}
version highly competitive.

Figure~\ref{refs} exhibits a desirable property: if we consider the
implementations that remove the incidental overheads that we
identified (and which otherwise dominate computation costs), all
language versions exhibit parallel scaling.  Observe the parallel
speedups in the case of \java{}, \scala{}-strict, \fs{}, and \cs{} on
Windows.

One final remark is on the choice of using the C2 JIT compiler only (by using
the -XX:-TieredCompilation flag). In both \scala{} and \java{} tests, using
tiered compilation degraded the performance in the majority of our
benchmarks. Concretely, for the \scala{} tests, tiered compilation had only a
minor positive effect on the \textbf{sum} tests and an approximately 10\%
performance degradation in all other cases. Regarding the \java{} cases, all
tests, apart from the sequential and parallel versions of the \textbf{refs}
benchmark, presented performance degradation.

\section{Future Work}

Several possibilities for further work arise. Our benchmark suite
can be enhanced with more complex microbenchmarks to
capture the case of streams that include a variable number
of successive combinators, such as \sv{filter}s. Additionally, an interesting
followup would be to examine how measurements are affected as a
function of the number of processors. Regarding standard stream APIs, \cs{},
\fs{} and \scala{} seem to use external iteration while \java{} uses internal iteration.
Thus an interesting direction is to implement internal iterator-based streaming
APIs for the aforementioned languages. Finally, LinqOptimizer demonstrated how,
by leveraging the LINQ Expression tree API, optimized queries can be obtained,
while ScalaBlitz employed macros for compile-time optimizations. \java{} can
benefit from an optimizing framework. As \java{} can have access to the internal
compiler API, a very promising direction to explore is the design and
development of an optimizing framework, designed as a \sv{javac} plugin.

\section{Conclusions}

In this work, we evaluated the combined cost of lambdas and stream
APIs in four different multiparadigm languages running on two
different runtime platforms. We used benchmarks expressed with the
closest comparable datatypes that each language offers in order to
preserve semantic equivalence. Our benchmarks constitute a fine
grained set. Each benchmark builds upon the previous one in terms of
complexity. Additionally we run all benchmarks on both Windows and
Linux. Our results clearly show the benefit of advanced implementation
techniques in \java{}, but also the performance advantage of
optimizing frameworks that can radically transform streaming
pipelines.

\section*{Acknowledgments} We would like to thank Aleksey Shipilev, Paul Sandoz,
Brian Goetz, Alex Buckley, Doug Lea, and the ScalaBlitz developers,
Aleksandar Prokopec and Dmitry Petrashko, for their valuable comments
that helped strengthen this study. We gratefully acknowledge funding
by the European Union under a Marie Curie International Reintegration
Grant (\textsc{PaDecl}) and a European Research Council
Starting/Consolidator grant (\textsc{Spade}); and by the Greek
Secretariat for Research and Technology under an Excellence (Aristeia)
award (\textsc{Morph-PL}.)

\bibliographystyle{abbrvnat}
% \bibliography{icooolps14}

\end{document}